\documentclass[10pt]{amsart}

\usepackage{lipsum}
\usepackage{amsfonts,bm,amssymb,amsmath}
\usepackage{graphicx}
\usepackage{epstopdf}
\usepackage[caption=false]{subfig}
\usepackage{pgfplots}
\usepackage{algorithmic}
\usepackage{amsopn}
\usepackage{amsrefs}
\usepackage{float}
\usepackage{caption}

\usepackage{hyperref}
\usepackage{cleveref}

\usepackage{color}
\usepackage{diagbox}

\newtheorem{remark}{Remark}[section]

\hypersetup{
	colorlinks,
	linkcolor={red!50!black},
	citecolor={blue!50!black},
	urlcolor={blue!80!black}
}

\numberwithin{equation}{section}

\definecolor{newcolor1}{rgb}{.8,.349,.1}
\colorlet{bblue}{blue!50!black}
\crefformat{equation}{(#2#1#3)}
\crefmultiformat{equation}{(#2#1#3)}{ and~(#2#1#3)}{, (#2#1#3)}{ and~(#2#1#3)}

\crefformat{figure}{Figure~#2#1#3}
\crefmultiformat{figure}{Figures~ #2#1#3}{ and~#2#1#3}{, (#2#1#3)}{ and~(#2#1#3)}
\crefformat{table}{Table~#2#1#3}

\def\e{\mbox{\boldmath $e$}}
\def\f{\mbox{\boldmath $f$}}

\def\g{\mbox{\boldmath $g$}}
\def\h{\mbox{\boldmath $h$}}
\def\m{\mbox{\boldmath $m$}}

\def\x{\mbox{\boldmath $x$}}
\def\y{\mbox{\boldmath $y$}}

\def\0{\mbox{\boldmath $0$}}

\begin{document}

\title[A second-order method for LLG equation]{A second-order numerical method for Landau-Lifshitz-Gilbert equation with large damping parameters}

\author[Y. Cai]{Yongyong Cai}
\address{School of Mathematical Sciences\\ Beijing Normal University\\ Beijing \\ China.}
\email{yongyong.cai@bnu.edu.cn}

\author[J. Chen]{Jingrun Chen}
\address{School of Mathematical Sciences\\ Soochow University\\ Suzhou\\ China.}
\email{jingrunchen@suda.edu.cn}

\author[C. Wang]{Cheng Wang}
\address{Mathematics Department\\ University of Massachusetts\\ North Dartmouth\\ MA 02747\\ USA.}
\email{cwang1@umassd.edu}
%\thanks{Support information for the second author.}

\author[C. Xie]{Changjian Xie}
\address{School of Mathematical Sciences\\ Soochow University\\ Suzhou\\ China.}
\email{20184007005@stu.suda.edu.cn}

%    General info

\date{\today}


\begin{abstract}
A second order accurate numerical scheme is proposed and implemented for the Landau-Lifshitz-Gilbert equation, which models magnetization dynamics in ferromagnetic materials, with large damping parameters. % is modeled by the Landau-Lifshitz-Gilbert equation and numerical methods are employed for real micromagnetics simulations. Semi-implicit schemes are popular since they are unconditionally stable on one hand and are efficient without the need of solving nonlinear systems of equations at each time step on the other hand. 
%Gauss-Seidel project method is one such scheme. It is unconditionally stable and only solves several linear systems of equations with symmetric and positive definite structures, which can be solved by fast solvers. However, its temporal accuracy is limited to be first-order. High-order semi-implicit schemes achieve high-order accuracy in time, but always need to solve linear systems of equations with nonsymmetric structures, in which case fast solvers are not available. In this work, we propose a second-order semi-implicit method to solve the Landau-Lifshitz-Gilbert equation with large damping parameters. 
The main advantages of this method are associated with the following features: (1) It only solves linear systems of equations with constant coefficients where fast solvers are available, so that the numerical efficiency has been greatly improved, in comparison with the existing Gauss-Seidel project method. (2) The second-order accuracy in time is achieved, and it is unconditionally stable for large damping parameters. Moreover, both the second-order accuracy and the great efficiency improvement will be verified by several numerical examples in the 1D and 3D simulations. In the presence of large damping parameters, it is observed that this method is unconditionally stable and finds physically reasonable structures while many existing methods have failed. For the domain wall dynamics, the linear dependence of wall velocity with respect to the damping parameter and the external magnetic field will be obtained through the reported simulations.
\end{abstract}

\maketitle

\section{Introduction}
Ferromagnetic materials are widely used for data storage due to the bi-stable states of the intrinsic magnetic order or magnetization. The dynamics of magnetization has been modeled by the Landau-Lifshitz-Gilbert (LLG) equation~\cite{Landau1935On,Gilbert:1955}. In particular, two terms are involved in the dynamics of the LLG equation: the gyromagnetic term, which is energetically conservative, and the damping term, which is energetically dissipative.

The damping term is important since it strongly affects the energy required and the speed at which a magnetic device operates. A recent experiment on a magnetic-semiconductor heterostructure~\cite{Zhang2020ExtremelyLM} has indicated that the Gilbert damping constant can be adjusted. At the microscopic level, the electron scattering, the itinerant electron relaxation~\cite{Heinrich1967TheIO}, and the phonon-magnon coupling~\cite{Suhl1998TheoryOT, Nan2020ElectricfieldCO} are responsible to the damping, which can be obtained from electronic structure calculations \cite{TangXia2017}. For the application purpose, tuning the damping parameter allows one to optimize the magneto-dynamic properties in the material, such as lowering the switching current and increasing the writing speed of magnetic memory devices~\cite{Wei2012MicromagneticsAR}. 

While most experiments have been devoted to small damping parameters \cite{Budhathoki2020LowGD,Lattery2018LowGD,Weber2019GilbertDO}, large damping effects are observed in \cite{GilbertKelly1955, Tanaka2014MicrowaveAssistedMR}. The magnetization switching time tends to be shorter in the presence of the large damping constant \cite{Tanaka2014MicrowaveAssistedMR}. Extremely large damping parameters ($\sim 9$) are presented in \cite{GilbertKelly1955}.

The LLG equation is a vectorial and nonlinear system with the fixed length of magnetization in a point-wise sense. Significant efforts have been devoted to design efficient and stable numerical methods for micromagnetics simulations; see~\cite{kruzik2006recent,cimrak2007survey} for reviews and references therein. Among the existing numerical works, semi-implicit schemes have been very popular since they avoid a complicated nonlinear solver while preserving the numerical stability; see~\cite{alouges2006convergence, gao2014optimal, Xie2018}, etc. In particular, the second-order accurate backward differentiation formula (BDF) scheme is constructed in \cite{Xie2018}, with a one-sided interpolation. In turn, a three-dimensional linear system needs to be solved at each time step,  with non-constant coefficients. Moreover, a theoretical analysis of the second order convergence estimate has been established in \cite{jingrun2019analysis} for such a BDF2 method. As another approach, a linearly implicit method in \cite{alouges2006convergence} introduces the tangent space to deal with the length constraint of magnetization, with the first-order temporal accuracy. As a further extension, high-order BDF schemes have been constructed and analyzed in a more recent work~\cite{Lubich2021}. An unconditionally unique solvability of the semi-implicit schemes has been proved in~\cite{jingrun2019analysis,Lubich2021}, while the convergence analysis has required a condition that the temporal step-size is proportional to the spatial grid-size. However, an obvious disadvantage has been observed for these semi-implicit schemes: the vectorial structure of the LLG equation leads to a non-symmetric linear system at each time step, which cannot be implemented by an FFT-based fast solver. In fact, the GMRES is often used, while its efficiency depends heavily on the temporal step-size and the spatial grid-size, and extensive numerical experiments have indicated much more expensive computational costs than  standard Poisson solvers~\cite{Xie2018}.

The Gauss-Seidel projection method (GSPM) is another popular set of numerical algorithms since only linear systems with constant coefficients need to be solved at each time step~\cite{wang2001gauss, Garcia2001Improved, Li2020TwoIG}. % and symmetric, positive definite (spd) structures need to be solved and fast solvers are available \cite{wang2001gauss,Garcia2001Improved, Li2020TwoIG}. 
This method is based on a combination of a Gauss-Seidel update of an implicit solver for the gyromagnetic term, the heat flow of the harmonic map, and a projection step to overcome the stiffness and the nonlinearity associated to the LLG equation. In this numerical approach, the implicit discretization is only applied to the scalar heat equation implicitly several times; therefore, the FFT-based fast solvers become available, due to the symmetric, positive definite (SPD) structures of the linear system. The original GSPM method~\cite{Wang2000} turns out to be  unstable for small damping parameters, while this issue has been resolved in \cite{Garcia2001Improved} with more updates of the stray field. Its numerical efficiency has been further improved by reducing the number of linear systems per time step \cite{Li2020TwoIG}. One little deficiency of GSPM is its first-order accuracy in time. 

  Meanwhile, in spite of these improvements, the GSPM method is computationally more expensive than the standard Poisson solver, because of the Gauss-Seidel iteration involved in the algorithm. An additional deficiency of the GSPM is its first-order accuracy in time. Moreover, most of the above-mentioned methods have been mainly focused on small damping parameters with the only exception in a theoretical work \cite{Lubich2021}. %since large damping parameters are needed in the proof of the convergence analysis. 
In other words, there has been no numerical method designed specifically for real micromagnetics simulations with large damping parameters. In this paper, we propose a second-order accurate numerical method to solve the LLG equation with large damping parameters, whose complexity is also comparable to solving the scalar heat equation. % implicitly and is similar to that of the improved GSPM \cite{Li2020TwoIG}. 
To achieve this goal, the LLG system is reformulated, in which the damping term is rewritten as a harmonic mapping flow. In turn, the constant-coefficient Laplacian part is treated by a standard BDF2 temporal discretization, and the associated dissipation will form the foundation of the numerical stability. Meanwhile, all the nonlinear parts, including both the gyromagnetic term and the remaining nonlinear expansions in the damping term, are computed by a fully explicit approximation, which is accomplished by a second order extrapolation formula. Because of this fully explicit treatment for the nonlinear parts, the resulting numerical scheme only requires a standard Poisson solver at each time step. This fact will greatly facilitate the computational efforts, since the FFT-based fast solver could be efficiently applied, due to the SPD structure of the linear system involved at each time step. In addition, the numerical stability has been demonstrated by extensive computational experiments, and these experiments has verified the idea that the dissipation property of the heat equation part would be able to ensure the numerical stability of the nonlinear parts, with large damping parameters.

%The temporal accuracy of the proposed method is second-order, which is comparable to other semi-implicit schemes. Therefore, our method combines the advantages of both the GSPM and the semi-implicit projection methods (SIPM) such as BDF2. This method is proved to be second-order accurate in time under conditions that the temporal step-size is proportional to the spatial grid-size and the damping parameter is greater than $3$ \cite{Xie2021EElinear}.

The rest of this paper is organized as follows. In \cref{sec: numerical scheme}, the micromagnetics model is reviewed, and the numerical method is proposed, as well as its comparison with the GSPM and the semi-implicit projection method (SIPM). Subsequently, the numerical results are presented in \cref{sec:experiments}, including the temporal and spatial accuracy check in both the 1D and 3D computations, the numerical efficiency investigation (in comparison with the GSPM and SIPM algorithms), the stability study with respect to the damping parameter, and the dependence of domain wall velocity on the damping parameter and the external magnetic field. Finally, some concluding remarks are made in \cref{sec:conclusions}.    

\section{The physical model and the numerical method}
\label{sec: numerical scheme}

\subsection{Landau-Lifshitz-Gilbert equation}

The LLG equation describes the dynamics of magnetization which consists of the gyromagnetic term and the damping term~\cite{Landau1935On,Brown1963micromagnetics}. In the nondimensionalized form, this equation reads as
\begin{align}\label{c1-large}
{\m}_t =-{\m}\times{\bm h}_{\text{eff}}-\alpha{\m}\times({\m}\times{\bm h}_{\text{eff}})
\end{align}
with the homogeneous Neumann boundary condition
\begin{equation}\label{boundary-large}
\frac{\partial{\m}}{\partial {\bm \nu}}\Big|_{\partial \Omega}=0,
\end{equation}
where $\Omega$ is a bounded domain occupied by the ferromagnetic material and $\bm \nu$ is unit outward normal vector along $\partial \Omega$. 

In more details, the magnetization ${\m}\,:\,\Omega\subset\mathbb{R}^d\to \mathbb{R}^3,d=1,2,3 $ is a three-dimensional vector field with a pointwise constraint $|\m|=1$. The first term on the right-hand side in \cref{c1-large} is the gyromagnetic term and the second term stands for the damping term, with $\alpha>0$ being the dimensionless damping coefficient.

The effective field ${\bm h}_{\text{eff}}$ is obtained by taking the variation of the Gibbs free energy of the magnetic body with respect to $\m$. The free energy includes the exchange energy, the anisotropy energy, the magnetostatic energy, and the Zeeman energy:  
\begin{equation}\label{LL-Energy}
F[\m] = \frac {\mu_0 M_s^2}{2} \left\{\int_\Omega \left( \epsilon|\nabla\m|^2 +
q\left(m_2^2 + m_3^2\right)
-2\h_e\cdot\m - \h_s\cdot\m \right)\mathrm{d}\x \right\} . 
\end{equation}
Therefore, the effective field includes the exchange field, the anisotropy field, the stray field $\h_s$, and the external field $\h_e$. For a uniaxial material, it is clear that 
\begin{align}
{\bm h}_{\text{eff}} =\epsilon\Delta\m-q(m_2\e_2+m_3\e_3)+\h_s+\h_e,
\end{align}
where the dimensionless parameters become $\epsilon=C_{ex}/(\mu_0 M_s^2L^2)$ and $q=K_u/(\mu_0 M_s^2)$ with $L$ the diameter of the ferromagnetic body and $\mu_0$ the permeability of vacuum. The unit vectors are given by ${\bm e}_2=(0,1,0)$, ${\bm e}_3=(0,0,1)$, and $\Delta$ denotes the standard Laplacian operator. For the Permalloy, an alloy of Nickel ($80\%$) and Iron ($20\%$), typical values of the physical parameters are given by: the exchange constant $C_{ex}=1.3\times 10^{-11}\,\textrm{J/m}$, the anisotropy constant $K_u = 100\, \textrm{J/}\textrm{m}^3$, the saturation magnetization constant $M_s = 8.0\times 10^{5}\,\textrm{A/m}$. The stray field takes the form
\begin{align}\label{eqn:div}
	{\h}_{\text{s}}=\frac{1}{4\pi}\nabla \int_{\Omega} \nabla\left( \frac{1}{|\x-\y|}\right)\cdot {\bm m}({\bm y})\,d{\bm y}.
\end{align}
If $\Omega$ is a rectangular domain, the evaluation of \eqref{eqn:div} can be efficiently done by the Fast Fourier Transform (FFT) \cite{Wang2000}.

For brevity, the following source term is defined 
\begin{align}\label{eq-4}
\f=-Q(m_2\e_2+m_3\e_3)+\h_s+\h_e.
\end{align}
and the original PDE system \cref{c1-large} could be rewritten as
\begin{align}\label{eq-5}
\m_t=-\m\times(\epsilon\Delta\m+\f)-\alpha\m\times\m\times(\epsilon\Delta\m+\f).
\end{align}
Thanks to point-wise identity $|\m|=1$, we obtain an equivalent form: 
\begin{equation}\label{eq-model}
\m_t=\alpha  (\epsilon\Delta\m+\f)+\alpha \left(\epsilon |\nabla \m|^2 -\m \cdot\f \right)\m-\m\times(\epsilon\Delta\m+\f).
\end{equation}
In particular, it is noticed that the damping term is rewritten as a harmonic mapping flow, which contains a constant-coefficient Laplacian diffusion term. This fact will greatly improve the numerical stability of the proposed scheme.  

For the numerical description, we first introduce some notations for discretization and numerical approximation. %First, the GSPM and the SIPM numerical methods need to be reviewed, which could be used for the later comparison. 
Denote the temporal step-size by $k$, and $t^n=nk$, $n\leq \left\lfloor\frac{T}{k}\right\rfloor$ with $T$ the final time. The spatial mesh-size is given by $h_x=h_y=h_z=h=1/N$, and $\m_{i,j,\ell}^n$ stands for the magnetization at time step $t^n$, evaluated at the spatial location $(x_{i-\frac12},y_{j-\frac12},z_{\ell-\frac12})$ with $x_{i-\frac12}=\left(i-\frac12\right)h_x$, $y_{j-\frac12}=\left(j-\frac12\right)h_y$ and $z_{\ell-\frac12}=\left(\ell-\frac12\right)h_z$ ($0\leq i,j,\ell\leq N+1$). In addition, a third order extrapolation formula is used to approximate the homogeneous Neumann boundary condition. For example, such a formula near the boundary along the $z$ direction is given by 
\begin{align*}
\m_{i,j,1}=\m_{i,j,0},\quad \m_{i,j,N+1}=\m_{i,j,N}.
\end{align*}
The boundary extrapolation along other boundary sections can be similarly made.

The standard second-order centered difference applied to $\Delta \m$ results in
\begin{align*}
\Delta_h \m_{i,j,k} &=\frac{\m_{i+1,j,k}-2\m_{i,j,k}+\m_{i-1,j,k}}{h_x^2}\\
&+\frac{\m_{i,j+1,k}-2\m_{i,j,k}+\m_{i,j-1,k}}{h_y^2}\nonumber\\
&+\frac{\m_{i,j,k+1}-2\m_{i,j,k}+\m_{i,j,k-1}}{h_z^2},\nonumber
\end{align*}
and the discrete gradient operator $\nabla_h \m$ with $\m=(u, v, w)^T$ reads as
\begin{align*}
\nabla_h\m_{i,j,k} = \begin{bmatrix}
\frac{u_{i+1,j,k}-u_{i-1,j,k}}{h_x}&\frac{v_{i+1,j,k}-v_{i-1,j,k}}{h_x}&\frac{w_{i+1,j,k}-w_{i-1,j,k}}{h_x}\\
\frac{u_{i,j+1,k}-u_{i,j-1,k}}{h_y}&\frac{v_{i,j+1,k}-v_{i,j-1,k}}{h_y}&\frac{w_{i,j+1,k}-w_{i,j-1,k}}{h_y}\\
\frac{u_{i,j,k+1}-u_{i,j,k-1}}{h_z}&\frac{v_{i,j,k+1}-v_{i,j,k-1}}{h_z}&\frac{w_{i,j,k+1}-w_{i,j,k-1}}{h_z}
\end{bmatrix}.
\end{align*}

Subsequently, the GSPM and the SIPM numerical methods need to be reviewed, which could be used for the later comparison. 

\subsection{The Gauss-Seidel projection method}

The GSPM is based on a combination of a Gauss-Seidel update of an implicit solver for the gyromagnetic term, the heat flow of the harmonic map, and a projection step. It only requires a series of heat equation solvers with constant coefficients; as a result, the FFT-based fast solvers could be easily applied. This method is first-order in time and second-order in space. Below is the detailed outline of the GSPM method in \cite{Garcia2001Improved}.
\begin{enumerate}
	\item[Step 1. ] Implicit Gauss-Seidel:
	\begin{align}
	g_i^n &= (I-\epsilon \Delta t\Delta_h)^{-1}(m_i^n+\Delta tf_i^n),\ \ i=2,3, \nonumber \\
	g_i^{*} &= (I-\epsilon \Delta t\Delta_h)^{-1}(m_i^{*}+\Delta tf_i^{*}),\ \ i=1,2,  \label{eq-17}
	\end{align}
	{\begin{equation}
		\begin{pmatrix}
		m_1^{*}\\
		m_2^{*}\\
		m_3^{*}
		\end{pmatrix}
		=
		\begin{pmatrix}
		m_1^n+(g_2^nm_3^n-g_3^nm_2^n)\\
		m_2^n+(g_3^nm_1^{*}-g_1^{*}m_3^n)\\
		m_3^n+(g_1^{*}m_2^{*}-g_2^{*}m_1^{*})
		\end{pmatrix}.
		\end{equation}}
	\item[Step 2.] Heat flow without constraints:
	\begin{equation}\label{eq-19}
	{\bm f}^{*}=-Q(m_2^{*}{\bm e}_2+m_3^{*}{\bm e}_3)+{\bm h}_s^{*}+{\bm h}_e,
	\end{equation}
	
	{\begin{equation}
		\begin{pmatrix}
		m_1^{**}\\
		m_2^{**}\\
		m_3^{**}
		\end{pmatrix}
		=
		\begin{pmatrix}
		m_1^{*}+\alpha \Delta t (\epsilon \Delta_hm_1^{**}+f_1^{*})\\
		m_2^{*}+\alpha \Delta t (\epsilon \Delta_hm_2^{**}+f_2^{*})\\
		m_3^{*}+\alpha \Delta t (\epsilon \Delta_hm_3^{**}+f_3^{*})
		\end{pmatrix}.
		\end{equation}}
	\item[Step 3.] Projection onto $S^2$:
	{\begin{equation}
		\begin{pmatrix}
		m_1^{n+1}\\
		m_2^{n+1}\\
		m_3^{n+1}
		\end{pmatrix}
		=
		\frac{1}{|m^{**}|}\begin{pmatrix}
		m_1^{**}\\
		m_2^{**}\\
		m_3^{**}
		\end{pmatrix}.
		\end{equation}}
\end{enumerate}
Here $\m^*$ denotes the intermediate values of $\m$, and stray fields $\h_s^n$ and $\h_s^*$ are evaluated at $\m^{n}$ and $\m^{*}$, respectively. 

\begin{remark}
Two improved versions of the GSPM have been studied in \cite{Li2020TwoIG}, which turn out to be more efficient than the original GSPM. Meanwhile, it is found that both improved versions become unstable when $\alpha>1$, while the original GSPM (outlined above) is stable even when $\alpha\leq10$. Therefore, we shall use the original GSPM in \cite{Garcia2001Improved} for the numerical comparison in this work.
\end{remark}

\subsection{Semi-implicit projection method}
The SIPM has been outlined in \cite{Xie2018,jingrun2019analysis}. This method is based on the second-order BDF temporal discretization, combined with an explicit extrapolation. It is found that SIPM is unconditionally stable and is second-order accurate in both space and time. The algorithmic details are given as follows.
\begin{equation}\label{sipm}
\left\{ 
\begin{aligned}
&\frac{\frac32 {\tilde{\m}}_h^{n+2} - 2 {\m}_h^{n+1} + \frac12 {\m}_h^n}{k}
=  - \hat{\m}_h^{n+2} \times\big(\epsilon \Delta_h\tilde{\m}_h^{n+2} +\hat{\f}_h^{n+2} \big) \\
&\quad - \alpha \hat{\m}_h^{n+2} \times \left(\hat{\m}_h^{n+2}\times(\epsilon \Delta_h\tilde{\m}_h^{n+2} +\hat{\f}_h^{n+2} ) \right), \\
& \qquad\qquad\qquad\qquad\quad \m_h^{n+2} = \frac{\tilde{\m}_h^{n+2}}{ |\tilde{\m}_h^{n+2}| },
\end{aligned}
\right.
\end{equation} 
where $\tilde{\m}_h^{n+2}$ is an intermediate magnetization, and $\hat{\m}_h^{n+2}$, $\hat{\f}_h^{n+2}$ are given by the following extrapolation formula: 
\begin{align*}
\hat{\m}_h^{n+2} &=2{\m}_h^{n+1}-{\m}_h^n, \label{m_hat}\\
\hat{\f}_h^{n+2} &=2{\f}_h^{n+1}-{\f}_h^n,
\end{align*}
with $\f_h^{n}=-Q(m_2^n\e_2+m_3^n\e_3)+\h_s^n+\h_e^n$. The presence of cross product in the SIPM yields a linear system of equations with non-symmetric structure and variable coefficients. In turn, the GMRES solver has to be applied to implement this numerical system. The numerical evidence has revealed that, the convergence of GMRES solver becomes slower for larger temporal step-size $k$ or smaller spatial grid-size $h$, which makes the computation more challenging. 

\subsection{The proposed numerical method} \label{discretisations}

The SIPM in \eqref{sipm} treats both the gyromagentic and the damping terms in a semi-implicit way, i.e., $\Delta \m$ is computed implicitly, while the coefficient functions are updated by a second order accurate, explicit extrapolation formula. The strength of the gyromagnetic term is controlled by $\Delta\m +\f$ since the length of $\m$ is always $1$. Meanwhile, the strength of the damping term is controlled by the product of $\Delta\m +\f$ and the damping parameter $\alpha$. For small $\alpha$, say $\alpha \leq 1$, it is reasonable to treat both the gyromagentic and the damping terms semi-implicitly. However, for large $\alpha$, an alternate approach would be more reasonable, in which the whole gyromagentic term is computed by an explicit extrapolation, while the nonlinear parts in the damping term is also updated by an explicit formula, and only the constant-coefficient $\Delta \m$ part in the damping term is implicitly updated. This idea leads to the proposed numerical method. To further simplify the presentation, we start with \eqref{eq-model}, and the numerical algorithm is proposed as follows. %We only treat the heat flow of the harmonic map implicitly, while keep all the remaining terms explicitly. This strategy yields the following method.
\begin{equation}\label{proposed}
\left\{ 
\begin{aligned}
&\frac{\frac32 \tilde{\m}_h^{n+2} - 2 {\m}_h^{n+1} + \frac12 {\m}_h^n}{k}
=  - \hat{\m}_h^{n+2} \times \left(\epsilon \Delta_h \hat{{\m}}_h^{n+2} +\hat{\f}_h^{n+2}\right) \\
&\quad + \alpha \left(\epsilon \Delta_h \tilde{\m}_h^{n+2}+\hat{\f}_h^{n+2}\right)\\  
&\quad + \alpha \left(\epsilon | \nabla_h \hat{\m}_h^{n+2} |^2-\hat{\m}_h^{n+2}\cdot \hat{\f}_h^{n+2}\right) \hat{\m}_h^{n+2},\\ 
&\m_h^{n+2} = \frac{\tilde{\m}_h^{n+2}}{ |\tilde{\m}_h^{n+2}| } ,
\end{aligned}
\right.
\end{equation}
where
\begin{align*}
\hat{\m}_h^{n+2} &= 2 \m_h^{n+1} - \m_h^n,\\
\hat{\f}_h^{n+2} &= 2 \f_h^{n+1} - \f_h^n.
\end{align*}

\cref{tab-features} compares the proposed method, the GSPM and the SIPM in terms of number of unknowns, dimensional size, symmetry pattern, and availability of FFT-based fast solver of linear systems of equations, and the number of stray field updates. At the formal level, the proposed method is clearly superior to both the GSPM and the SIPM algorithms. In more details, this scheme will greatly improve the computational efficiency, since only three Poisson solvers are needed at each time step. Moreover, this numerical method preserves a second-order accuracy in both space and time. The numerical results in \cref{sec:experiments} will demonstrate that the proposed scheme provides a reliable and robust approach for micromagnetics simulations with high accuracy and efficiency in the regime of large damping parameters. 
\begin{table}[htbp]
	\begin{center}
		\caption{Comparison of the proposed method, the Gauss-Seidel projection method, and the semi-implicit projection method.}\label{tab-features}
		\begin{tabular}{cccc}
			\hline
			Property or number & Proposed method & GSPM & SIPM\\
			\hline
			Linear systems& \boldsymbol{$3$} & $7$ & $1$ \\
			Size & \boldsymbol{$N^3$} & $N^3$& $3N^3$ \\
			Symmetry& {\bf Yes}& Yes& No \\
			Fast Solver& {\bf Yes}& Yes& No \\
			Accuracy& \boldsymbol{$\mathcal{O}(k^2+h^2)$} & $\mathcal{O}(k+h^2)$ & $\mathcal{O}(k^2+h^2)$ \\
			Stray field updates & \boldsymbol{$1$} &$4$ &$1$ \\
			\hline
		\end{tabular}
	\end{center}
\end{table}

\begin{remark}
To kick start the proposed method, one can apply a first-order algorithm, such as the first-order BDF method, in the first time step. %treating the linear term $\alpha \epsilon \Delta \m$ implicitly, and keeping the remaining terms explicitly. 
An overall second-order accuracy is preserved in this approach.
\end{remark}

\section{Numerical experiments}
\label{sec:experiments}

In this section, we present a few numerical experiments with a sequence of damping parameters for the proposed method, the GSPM \cite{Garcia2001Improved} and the SIPM \cite{Xie2018}, with the accuracy, efficiency, and stability examined in details. Domain wall dynamics is studied and its velocity is recorded in terms of the damping parameter and the external magnetic field. 

\subsection{Accuracy and efficiency tests}

We set $\epsilon=1$ and $\f=0$ in \cref{eq-model} for convenience. The 1D exact solution is given by 
\begin{equation*}
	\m_e=\left(\cos(X)\sin t, \sin(X)\sin t, \cos t\right)^T,
\end{equation*}
and the corresponding exact solution in 3D becomes 
\begin{equation*}
\m_e=\left(\cos(XYZ)\sin t, \sin(XYZ)\sin t, \cos t\right)^T,
\end{equation*}
where $X=x^2(1-x)^2$, $Y=y^2(1-y)^2$, $Z=z^2(1-z)^2$. In fact, the above exact solutions satisfy \cref{eq-model} with the forcing term $\g=\partial_t \m_e-\alpha \Delta \m_e -\alpha |\nabla \m_e|^2+\m_e \times \Delta \m_e$, as well as the homogeneous Neumann boundary condition. 

For the temporal accuracy test in the 1D case, we fix the spatial resolution as $h=5D-4$, so that the spatial approximation error becomes negligible. The damping parameter is taken as $\alpha=10$, and the final time is set as $T=1$. In the 3D test for the temporal accuracy, 
due to the limitation of spatial resolution, we take a sequence of spatial and temporal mesh sizes: $k=h_x^2=h_y^2=h_z^2=h^2=1/N_0$ for the first-order method and $k=h_x=h_y=h_z=h=1/N_0$ for the second-order method, with the variation of $N_0$ indicated below. Similarly, the damping parameter is given by $\alpha=10$, while the final time $T$ is indicated below. In turn, the numerical errors are recorded in term of the temporal step-size $k$ in \cref{tab-1}. It is clear that the temporal accuracy orders of the proposed numerical method, the GSPM, and the SIPM are given by $2$, $1$, and $2$, respectively, in both the 1D and 3D computations. 

\begin{table}[htbp] 
	\centering
	{\caption{The numerical errors for the proposed method, the GSPM and the SIPM with $\alpha=10$ and $T=1$. Left: 1D with $h=5D-4$; Right: 3D with $k=h_x^2=h_y^2=h_z^2=h^2=1/N_0$ for GSPM and $k=h_x=h_y=h_z=h=1/N_0$ for the proposed method and SIPM, with $N_0$ specified in the table.}\label{tab-1} }{
		\subfloat[Proposed method]{\label{tab:floatrow:one}%	
			\begin{tabular}{cccc|cccc} 
				\hline	
				%		Proposed method & & & & & & & \\
				1D  & & {} & {} &3D &{} & {} &{} \\
				$k$ & $\|\cdot\|_{\infty}$ & $\|\cdot\|_{2}$ & $\|\cdot\|_{H^1}$ & 	$k=h$ & $\|\cdot\|_{\infty}$ & $\|\cdot\|_{2}$ & $\|\cdot\|_{H^1}$ \\
				%		$k$ & $\|\cdot\|_{\infty}$ & $\|\cdot\|_{2}$ & $\|\cdot\|_{H^1}$\\
				\hline
			4.0D-2& 4.459D-4& 5.226D-4& 5.588D-4& 1/20 & 6.171D-4 &4.240D-4 &4.246D-4 \\	
			2.0D-2 & 1.147D-4 &1.345D-4 & 1.436D-4 &		1/24 & 4.381D-4&3.010D-4 &3.014D-4 \\
			1.0D-2 & 2.899D-5 &3.402D-5 & 3.631D-5&	 1/28 & 3.268D-4&2.245D-4 & 2.248D-4\\
			5.0D-3 & 7.192D-6 & 8.529D-6& 9.119D-6 &	 1/32 & 2.531D-4 &1.739D-4 &1.741D-4\\
			2.5D-3 & 1.699D-6 & 2.321D-6 & 2.518D-6 & 	1/36& 2.017D-4&1.386D-4 & 1.387D-4\\
				order & 2.007 & 1.961 & 1.957 &{--}&1.902 &1.903 &1.903\\
				%		\hline \hline
				
				%		\hline \hline
				
				%		\hline \hline
				%			{} & {} & Semi-implicit& {} \\
				%	
				\hline
			\end{tabular}	
		}
		\qquad
		\subfloat[GSPM]{\label{tab:floatrow:two}
			\begin{tabular}{cccc|cccc} 
				\hline
				%		GSPM & & & & & & & \\
				1D & &  & {} & 3D & {} &  & {} \\
				$k$ & $\|\cdot\|_{\infty}$ & $\|\cdot\|_{2}$ & $\|\cdot\|_{H^1}$ & $k=h^2$ & $\|\cdot\|_{\infty}$ & $\|\cdot\|_{2}$ & $\|\cdot\|_{H^1}$\\
				\hline
				2.5D-3 & 2.796D-4&2.264D-4 & 1.445D-3& 1/36 & 4.194D-4 &2.683D-4 &2.815D-4 \\
				1.25D-3 &1.425D-4 & 1.174D-4&7.720D-4 & 1/64 & 2.388D-4& 1.399D-4&1.500D-4 \\
				6.25D-4 &7.170D-5 &5.940D-5 &4.026D-4 & 1/144 & 1.069D-4& 6.106D-5&6.736D-5  \\
				3.125D-4 & 3.591D-5 & 2.971D-5 & 2.069D-4 &1/256&6.021D-5 &3.442D-5 &3.860D-5 \\
				1.5625D-4& 1.799D-5 & 1.488D-5&1.054D-4 &1/400& 3.855D-5& 2.208D-5& 2.501D-5\\
				order & 0.991 & 0.984 &0.945 &{--}& 0.992&1.032 &1.000 \\
				%		\hline \hline
				
				%		\hline \hline
				
				%		\hline \hline
				%			{} & {} & Semi-implicit& {} \\
				%	
				\hline
			\end{tabular}
		}
		\qquad
		\subfloat[SIPM]{\label{tab:floatrow:three}
			\begin{tabular}{cccc|cccc} 
				\hline
				%	SIPM & & & & & & & \\
				1D  & & & {} & 3D & {} &  & {}\\
				$k$ & $\|\cdot\|_{\infty}$ & $\|\cdot\|_{2}$ & $\|\cdot\|_{H^1}$ & $k=h$ & $\|\cdot\|_{\infty}$ & $\|\cdot\|_{2}$ & $\|\cdot\|_{H^1}$\\
				\hline
				4.0D-2& 4.315D-4& 5.111D-4& 8.774D-4& 1/20 &6.170D-4 &4.240D-4 & 4.249D-4\\	
			2.0D-2 & 1.128D-4 &1.334D-4 &2.255D-4& 1/24 &4.380D-4 &3.010D-4 &3.016D-4 \\
				1.0D-2 & 2.872D-5& 3.399D-5& 5.706D-5 &	1/28 &3.268D-4 &2.245D-4 &2.251D-4 \\
			5.0D-3 &7.174D-6 &8.552D-6 &1.433D-5&	 1/32 &2.531D-4 &1.739D-4 &1.743D-4 \\
				2.5D-3 & 1.721D-6& 2.333D-6 & 3.784D-6 & 1/36& 2.017D-4& 1.386D-4&1.389D-4 \\
				order &1.991 &1.951 & 1.969 & {--}& 1.902&1.903 &1.902\\
				%		\hline \hline
				
				%		\hline \hline
				
				%		\hline \hline
				%			{} & {} & Semi-implicit& {} \\
				%	
				\hline
			\end{tabular}
	} }	
\end{table}

The spatial accuracy order is tested by fixing $k=1D-5$, $\alpha=10$, $T=1$ in 1D and $k=1D-3$, $\alpha=10$, $T=1$ in 3D. The numerical error is recorded in term of the spatial grid-size $h$ in \cref{tab-2}. Similarly, the presented results have indicated the second order spatial accuracy of all the numerical algorithms, including the proposed method, the GSPM, and the SIPM, respectively, in both the 1D and 3D computations. 

\begin{table}[htbp]
	\centering
	{\caption{The numerical errors of the proposed method, the GSPM and the SIPM with $\alpha=10$ and $T=1$. Left: 1D with $k=1D-5$; Right: 3D with $k=1D-3$.} \label{tab-2} }{
		\subfloat[Proposed method]{\label{tab:floatrow:1-S}
			\begin{tabular}{cccc|cccc}	
				\hline
				1D  & & & {} & 3D & & & \\
				$h$ & $\|\cdot\|_{\infty}$ &$\|\cdot\|_{2}$ &$\|\cdot\|_{H^1}$& $h$ & $\|\cdot\|_{\infty}$ & $\|\cdot\|_{2}$ & $\|\cdot\|_{H^1}$  \\
				\hline
				4.0D-2 & 7.388D-3& 7.392D-3&8.243D-3 & 1/2 & 4.261D-3 & 2.472D-3 & 2.472D-3\\
				2.0D-2 & 1.848D-3 & 1.848D-3 & 2.061D-3 & 1/4 & 9.822D-4 & 5.595D-4 & 5.753D-4\\
				1.0D-2 & 4.621D-4 &4.621D-4 & 5.153D-4 & 1/8 & 2.453D-4 &1.390D-4 &1.424D-4\\
				5.0D-3 & 1.155D-4 &1.155D-4 &1.288D-4 & 1/16 & 6.137D-5 & 3.471D-5 & 3.554D-5 \\
%				&  & &  & & & & \\
				order  & 2.000 & 2.000 &2.000 & {--} & 2.035& 2.047 &2.037  \\
				\hline
			\end{tabular}
		}	
		\qquad
		\subfloat[GSPM]{\label{tab:floatrow:2-S}
			\begin{tabular}{cccc|cccc}	
				\hline
				1D  & &  & {} & 3D & & & \\
				$h$ & $\|\cdot\|_{\infty}$ &$\|\cdot\|_{2}$ &$\|\cdot\|_{H^1}$ & $h$ & $\|\cdot\|_{\infty}$ &$\|\cdot\|_{2}$ &$\|\cdot\|_{H^1}$\\
				\hline
				4.0D-2 & 7.388D-3 & 7.392D-3 & 8.244D-3  & 1/2 &4.256D-3 & 2.470D-3 & 2.470D-3\\
				2.0D-2 & 1.848D-3 & 1.848D-3 &2.061D-3 & 1/4 & 9.810D-4& 5.589D-4 & 5.744D-4 \\
				1.0D-2 & 4.619D-4&4.622D-4 &5.158D-4 & 1/8 & 2.447D-4&1.388D-4 & 1.423D-4 \\
				5.0D-3 & 1.153D-4&1.156D-4 & 1.302D-4 & 1/16 &6.103D-5 & 3.468D-5 & 3.613D-5 \\
%				2.5D-3 &2.873D-5 &2.898D-5 &3.587D-5 & & & & \\
				order  & 2.000&2.000 & 1.995 & {--} & 2.037& 2.047 &2.030 \\
				\hline
			\end{tabular}
		}
		\qquad
		\subfloat[SIPM]{\label{tab:floatrow:3-S}
			\begin{tabular}{cccc|cccc}	
				\hline
				1D  & &  & & 3D & & & \\
				$h$ & $\|\cdot\|_{\infty}$ &$\|\cdot\|_{2}$ &$\|\cdot\|_{H^1}$ &$h$ & $\|\cdot\|_{\infty}$ &$\|\cdot\|_{2}$ &$\|\cdot\|_{H^1}$ \\
				\hline
				4.0D-2 &7.388D-3 & 7.392D-3& 8.243D-3 & 1/2 & 4.261D-3& 2.472D-3&2.472D-3\\
				2.0D-2 &1.848D-3 &1.848D-3 &2.061D-3 & 1/4 & 9.822D-4&5.595D-4 &5.753D-4 \\
				1.0D-2 & 4.621D-4& 4.621D-4& 5.153D-4 & 1/8 & 2.453D-4& 1.390D-4&1.424D-4\\
				5.0D-3 &1.155D-4 &1.155D-4 &1.288D-4 & 	1/16 & 6.137D-5& 3.471D-5& 3.554D-5 \\
%				2.5D-3 & 2.888D-5 &2.888D-5 &3.221D-5 & & & & \\
				order  &2.000 &2.000 &2.000& {--} &2.035 & 2.047&2.037 \\
				\hline
			\end{tabular} 
		}
	}
\end{table}

\begin{figure}[htbp]
	\centering
	\subfloat[Varying $k$ in 1D up to $T=1$ ]{\label{cputime_1D}\includegraphics[width=2.5in]{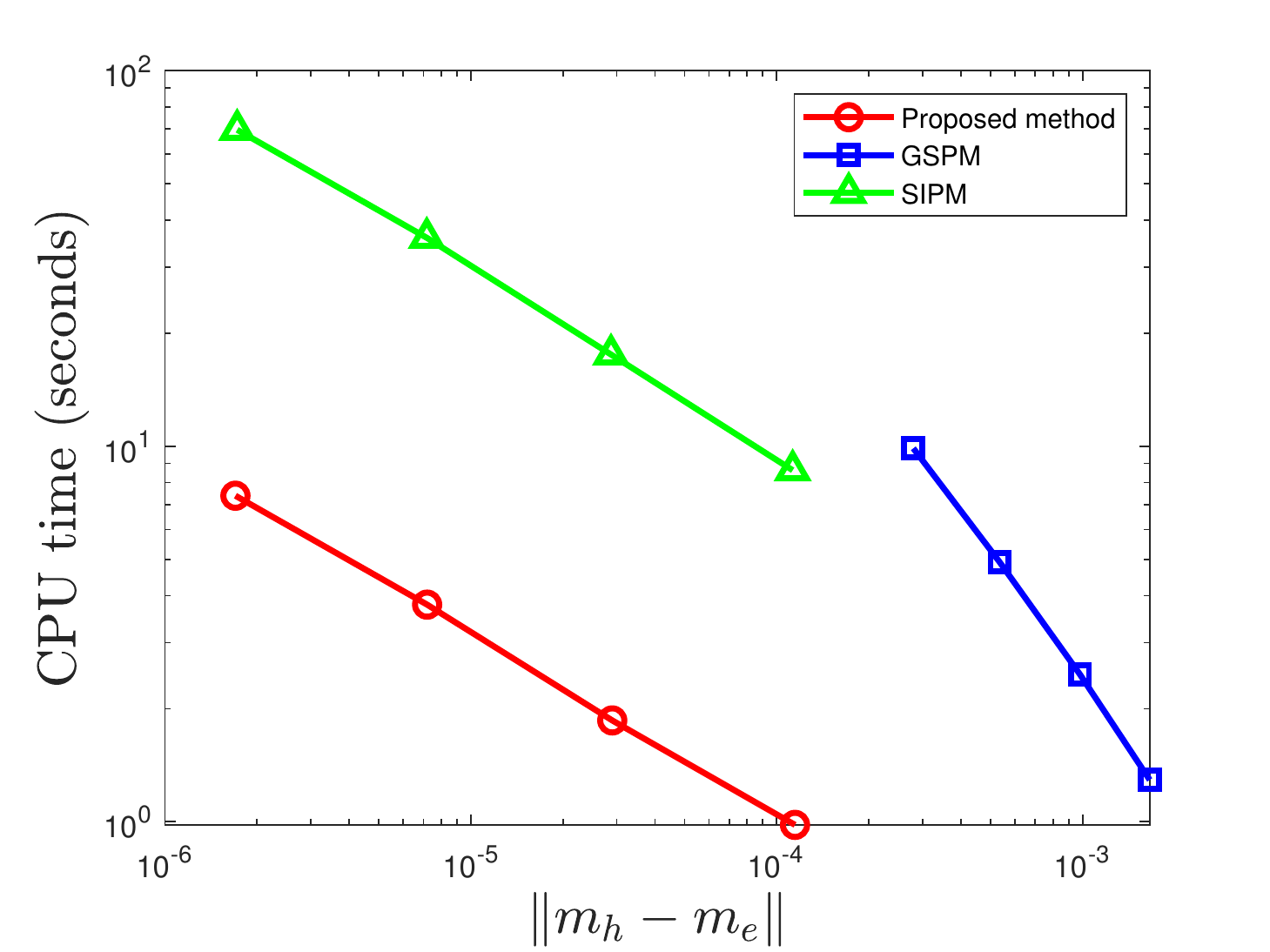}}
	\subfloat[Varying $k$ in 3D up to $T=0.1$]{\label{cputime_3D}\includegraphics[width=2.5in]{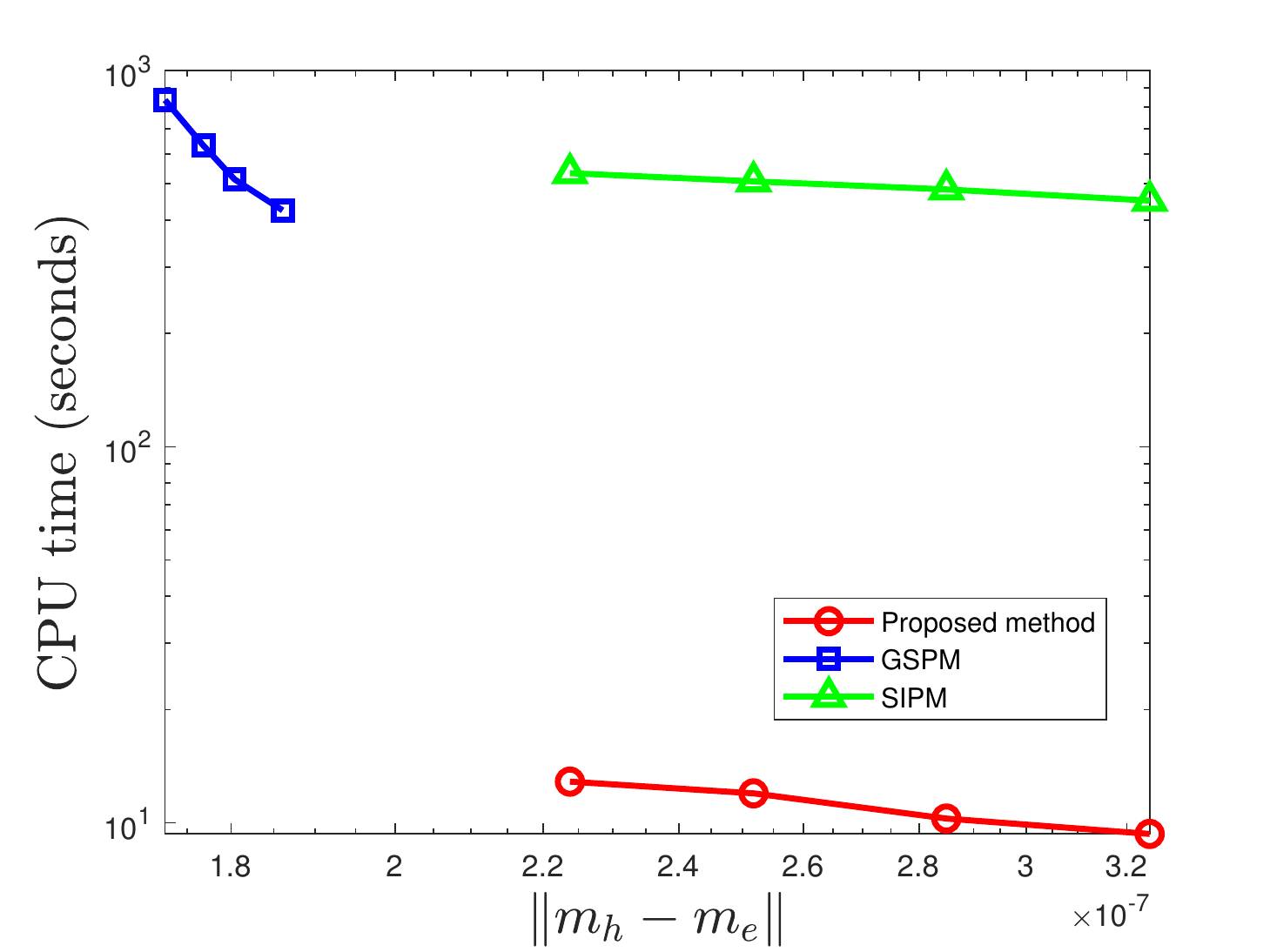}}
	\hspace{0.1in}
	\subfloat[Varying $h$ in 1D up to $T=1$]{\label{cputime_1D_space}\includegraphics[width=2.5in]{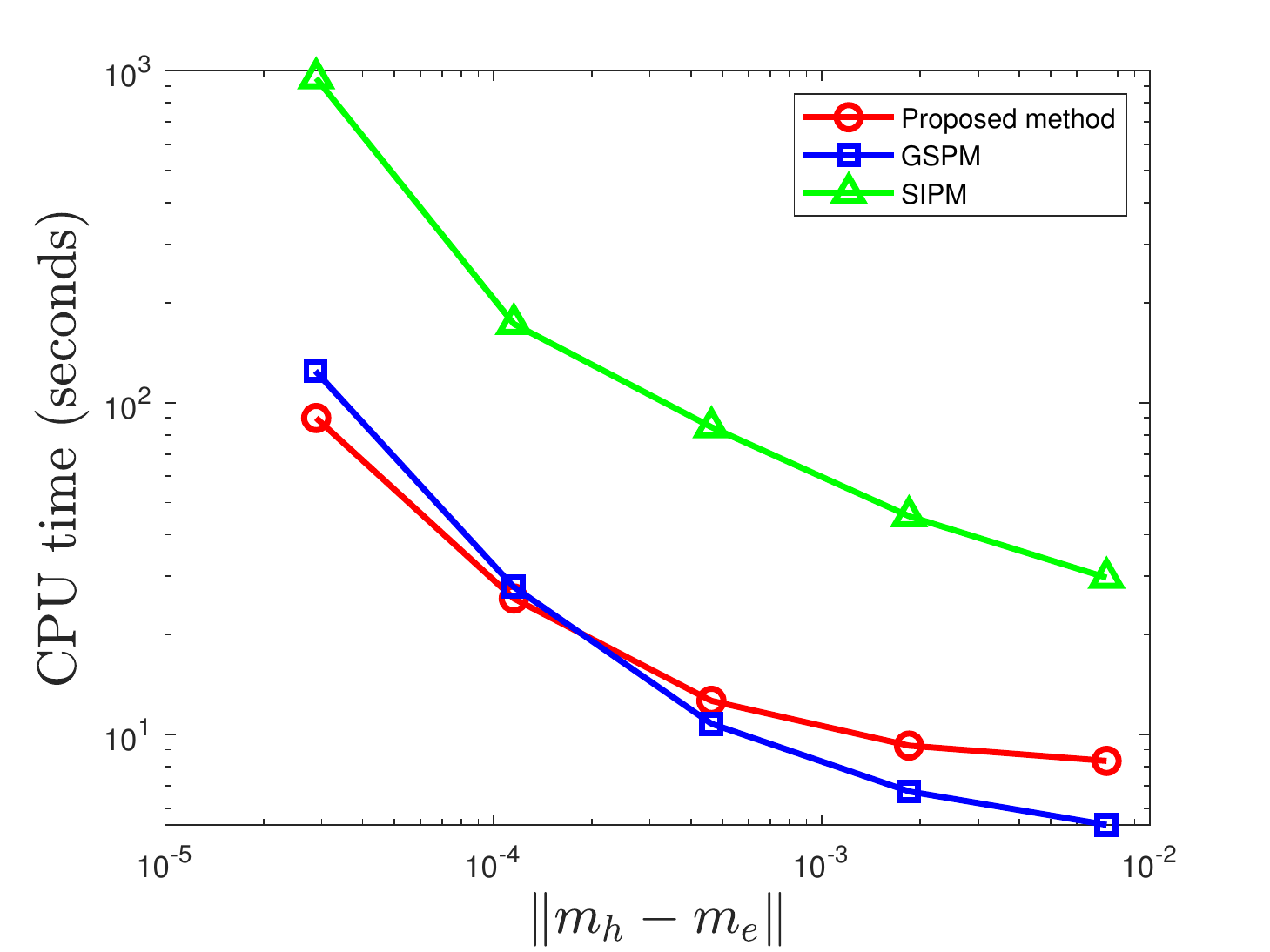}}
	\subfloat[Varying $h$ in 3D up to $T=1$]{\label{cputime_3D_space}\includegraphics[width=2.5in]{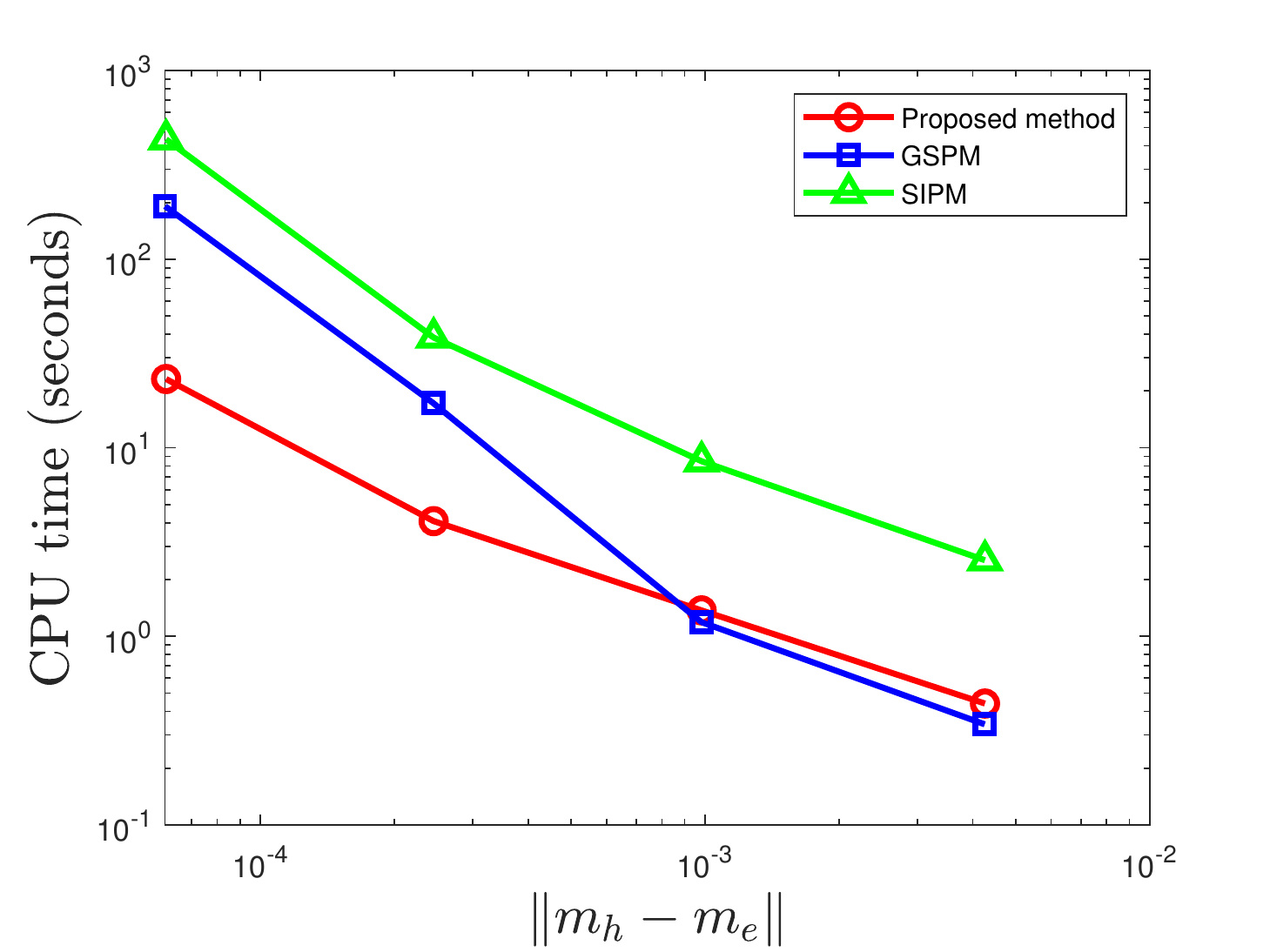}}
	\caption{CPU time needed to achieve the desired numerical accuracy, for the proposed method, the GSPM and the SIPM, in both the 1D and 3D computations. The CPU time is recorded as a function of the approximation error by varying $k$ or $h$ independently. CPU time with varying $k$: proposed method $<$ SIPM $<$ GSPM; CPU time with varying $h$: proposed method $\lessapprox$ GSPM $<$ SIPM.}\label{cputime}
\end{figure}

To make a comparison in terms of the numerical efficiency, we plot the CPU time (in seconds) vs. the error norm $\|\m_h-\m_e\|_{\infty}$. In details, the CPU time is recorded as a function of the approximation error in \cref{cputime_1D} in 1D and in \cref{cputime_3D} in 3D, with a variation of $k$ and a fixed value of $h$. Similar plots are also displayed in \cref{cputime_1D_space} in 1D and \cref{cputime_3D_space} in 3D, with a variation of $h$ and a fixed value of $k$. In the case of a fixed spatial resolution $h$, the proposed method is significantly more efficient than the GSPM and the SIPM in both the 1D and 3D computations. The SIPM is slightly more efficient than the GSPM, while such an advantage depends on the performance of GMRES, which may vary for different values of $k$ and $h$. In the case of a fixed time step size $k$, the proposed method is slightly more efficient than the GSPM, in both the 1D and 3D computations, and the GSPM is more efficient than the SIPM.  

\subsection{Stability test with large damping parameters}
\begin{figure}[htbp]
	\centering
%	\subfloat{\label{BDF2_alpha_40_mag}\includegraphics[width=1.3in]{BDF2_alpha_40_mag_v3_v2.png}}
%	\subfloat{\label{BDF2_alpha_10_mag}\includegraphics[width=1.3in]{BDF2_alpha_10_mag_v2_v1.png}}
%	\subfloat{\label{BDF2_alpha_1D-2_mag}\includegraphics[width=1.3in]{BDF2_alpha_1D-2_mag_v2_v1.png}}
%	\hspace{0.1in}
	\subfloat{\label{BDF2_alpha_40_ang}\includegraphics[width=1.3in]{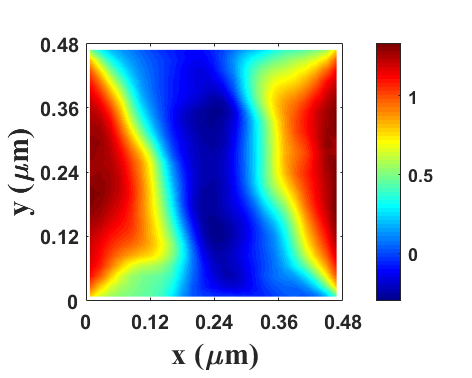}}
	\subfloat{\label{BDF2_alpha_10_ang}\includegraphics[width=1.3in]{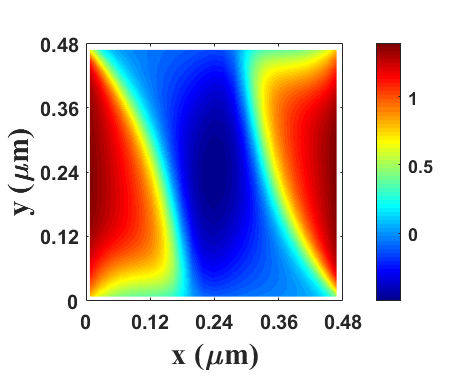}}
	\subfloat{\label{BDF2_alpha_1D-2_ang}\includegraphics[width=1.3in]{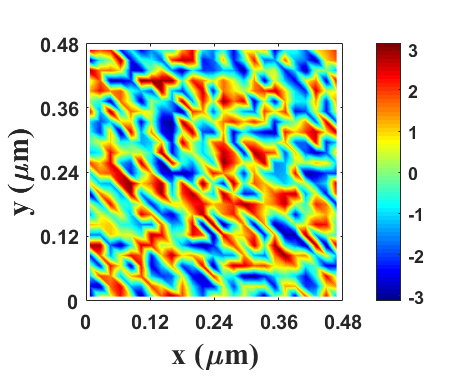}}
	\hspace{0.1in}
%	\subfloat{\label{improvedGSPM_alpha_40_mag}\includegraphics[width=1.3in]{improved_GSPM_alpha_40_mag_v2.png}}
%	\subfloat{\label{improvedGSPM_alpha_10_mag}\includegraphics[width=1.3in]{improved_GSPM_alpha_10_mag_v2.png}}
%	\subfloat{\label{improvedGSPM_alpha_1D-2_mag}\includegraphics[width=1.3in]{Improved_GSPM_alpha_1D-2_mag_v2.png}}
%	\hspace{0.1in}
	\subfloat{\label{improvedGSPM_alpha_40_ang}\includegraphics[width=1.3in]{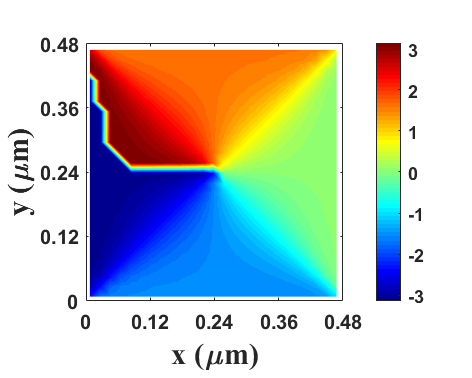}}
	\subfloat{\label{improvedGSPM_alpha_10_ang}\includegraphics[width=1.3in]{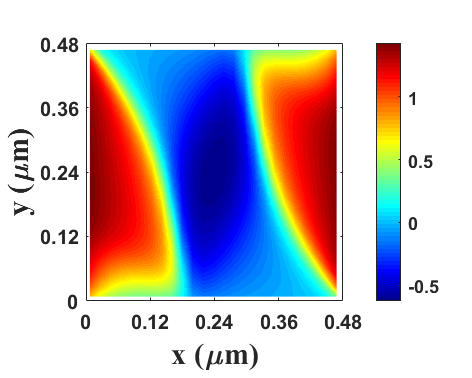}}
	\subfloat{\label{improvedGSPM_alpha_1D-2_ang}\includegraphics[width=1.3in]{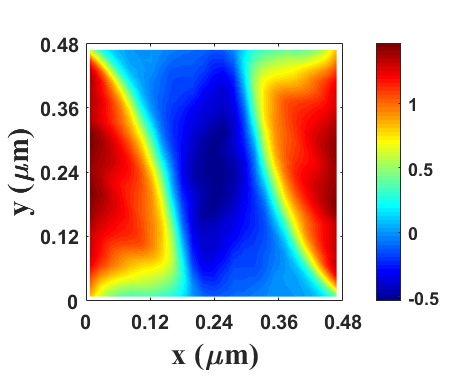}}
	\hspace{0.1in}
%	\subfloat{\label{semiBDF2_alpha_40_mag}\includegraphics[width=1.3in]{semi_implicit_BDF2_alpha_40_mag_v1.png}}
%	\subfloat{\label{semiBDF2_alpha_10_mag}\includegraphics[width=1.3in]{semiBDF2_alpha_10_mag_v1.png}}
%	\subfloat{\label{semiBDF2_alpha_1D-2_mag}\includegraphics[width=1.3in]{semiBDF2_alpha_1D-2_mag_v1.png}}
%	\hspace{0.1in}
	\subfloat{\label{semiBDF2_alpha_40_ang}\includegraphics[width=1.3in]{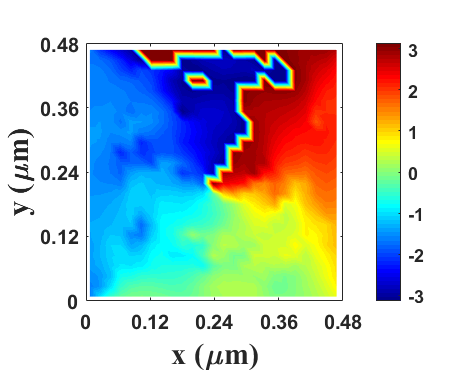}}
	\subfloat{\label{semiBDF2_alpha_10_ang}\includegraphics[width=1.3in]{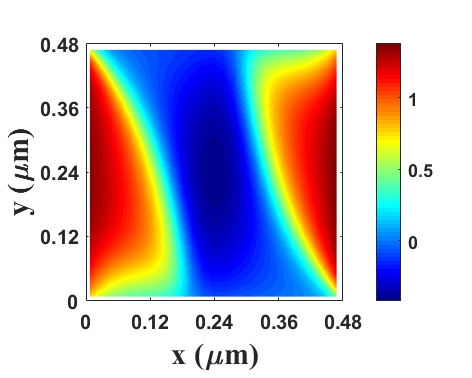}}
	\subfloat{\label{semiBDF2_alpha_1D-2_ang}\includegraphics[width=1.3in]{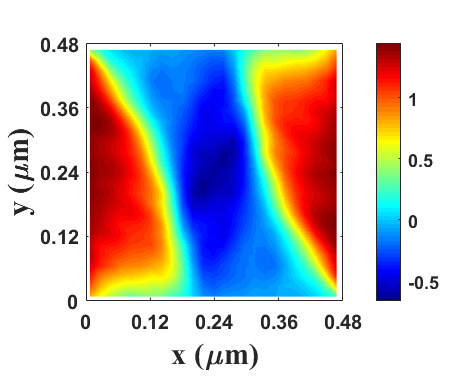}}
	\caption{Stable structures in the absence of magnetic field at $2\,$ns when $\alpha=0.01,10,40$. The color denotes the angle between the first two components of the magnetization vector. Top: Proposed method; Middle: GSPM; Bottom: SIPM. Left: $\alpha=40$; Middle: $\alpha=10$; Right: $\alpha=0.01$. }\label{BDF2_GSPM_alpha}
\end{figure}

To check the numerical stability of these three methods in the practical simulations of micromagnetics with large damping parameters, we consider a thin film of size $480\times480\times20\,\textrm{nm}^3$ with grid points $100\times100\times4$. The temporal step-size is taken as $k=1\,$ps. A uniform state along the $x$ direction is set to be the initial magnetization and the external magnetic field is set to be $0$. Three different damping parameters, $\alpha=0.01,10,40$, are tested with stable magnetization profiles shown in \cref{BDF2_GSPM_alpha}. In particular, the following observations are made. 
\begin{itemize}
	\item The proposed method is the only one that is stable for very large damping parameters;
	\item All three methods are stable for moderately large $\alpha$;
	\item The proposed method is the only one that is unstable for small $\alpha$.
\end{itemize}
%It is proven in \cite{Xie2021EElinear} that the rigorous convergence analysis 
In fact, a preliminary theoretical analysis reveals that, an optimal rate convergence estimate of the proposed method could be theoretically justified for $\alpha>3$. Meanwhile, extensive numerical experiments have implied that $\alpha>1$ is sufficient to ensure the numerical stability in the practical computations.

\begin{figure}[htbp]
	\centering
	\subfloat[Proposed]{\label{energy_BDF2}\includegraphics[width=1.8in]{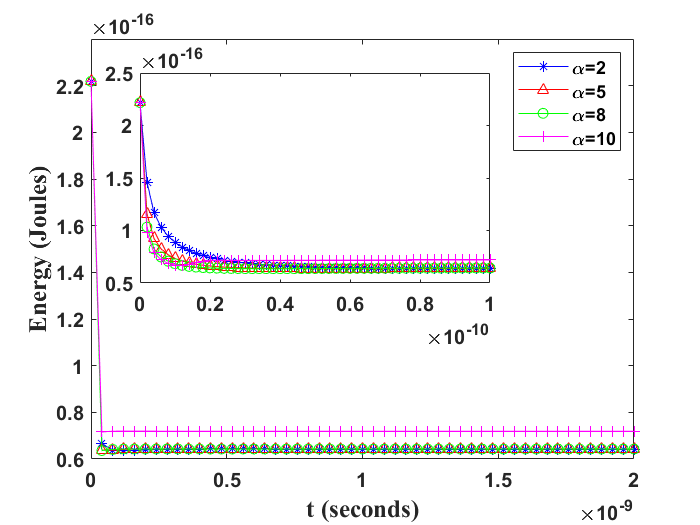}}
	\subfloat[GSPM]{\label{energy_GSPM_improved}\includegraphics[width=1.8in]{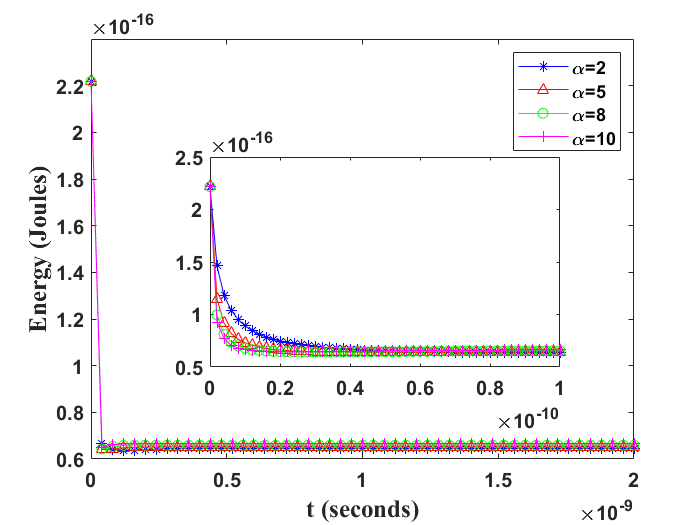}}
	\subfloat[SIPM]{\label{energy_semiBDF2}\includegraphics[width=1.8in]{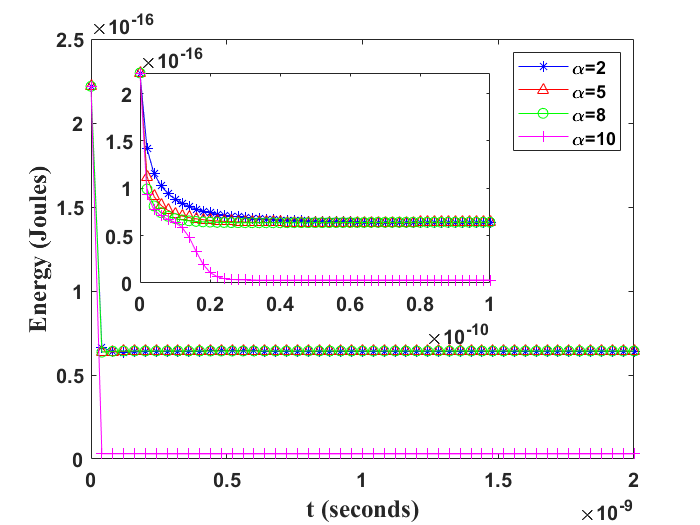}}
	\caption{Energy evolution curves of three numerical methods, with different damping constants, $\alpha=2,5,8,10$, up to $t=2\,$ns  in the absence of external magnetic field. Left: Proposed numerical method; Middle: GSPM; Right: SIPM. One common feature is that the energy dissipation rate is faster for larger $\alpha$, which is physically reasonable.}\label{energy_decay}
\end{figure}

Under the same setup outlined above, we investigate the energy dissipation of the proposed method, the GSPM, and the SIPM. The stable state is attainable at $t=2\,\textrm{ns}$, while the total energy is computed by \eqref{LL-Energy}. The energy evolution curves of different numerical methods with different damping parameters, $\alpha=2,5,8,10$, are displayed in \cref{energy_decay}. One common feature is that the energy dissipation rate turns out to be faster for larger $\alpha$, in all three schemes. Meanwhile, a theoretical derivation also reveals that the energy dissipation rate in the LLG equation \eqref{c1-large} depends on $\alpha$, and a larger $\alpha$ leads to a faster energy dissipation rate. Therefore, the numerical results generated by all these three numerical methods have made a nice agreement with the theoretical derivation. %provide the physically reasonable energy dissipation.

\begin{figure}[htbp]
	\centering
	\subfloat[$\alpha=2$]{\label{alpha_2}\includegraphics[width=2.5in]{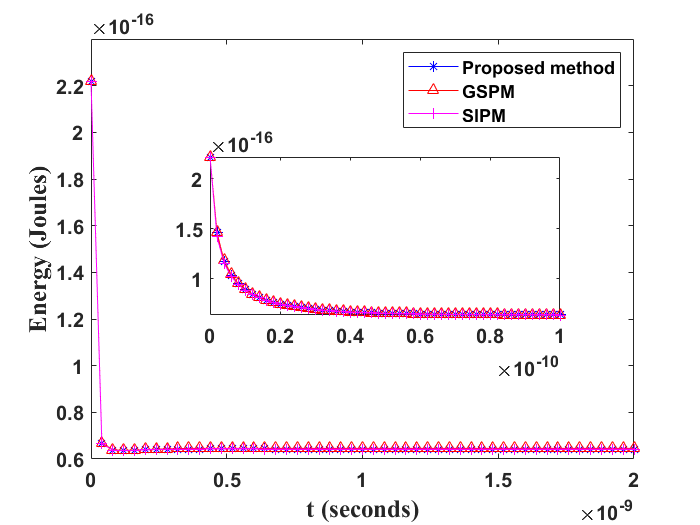}}
	\subfloat[$\alpha=5$]{\label{alpha_5}\includegraphics[width=2.5in]{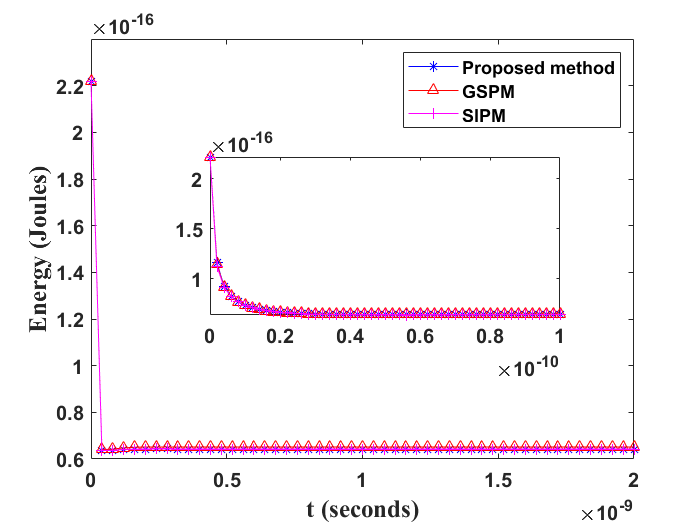}}
	\hspace{0.1in} 
	\subfloat[$\alpha=8$]{\label{alpha_8}\includegraphics[width=2.5in]{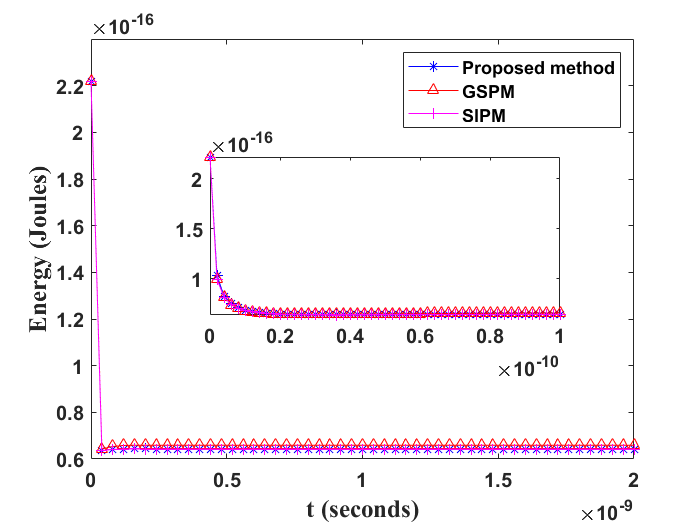}}
	\subfloat[$\alpha=10$]{\label{alpha_10}\includegraphics[width=2.5in]{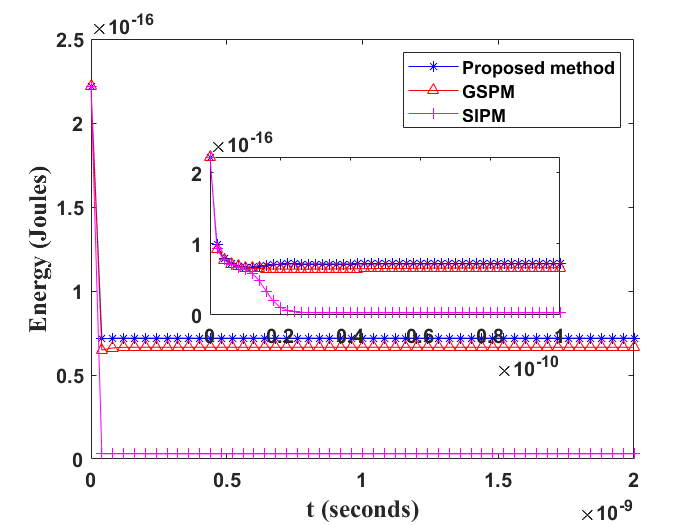}}
	\caption{Energy evolution curves in terms of time, for the numerical results created by three numerical methods up to $t=2\,$ns in the absence of external magnetic field for (a) $\alpha=2$, (b) $\alpha=5$, (c) $\alpha=8$, and (d) $\alpha=10$. The energy dissipation pattern of the proposed method is consistent with the other two methods for (a), (b), and (c), and the SIPM has a slightly different energy dissipation pattern from the other two methods for (d).}\label{energy_decay_alpha}
\end{figure}

Meanwhile, we choose the same sequence of values for $\alpha$, and display the energy evolution curves in terms of time up to $T=2\,$ns in \cref{energy_decay_alpha}. It is found that the proposed method have almost the same energy dissipation pattern with the other two methods for moderately large damping parameters $\alpha=2, 5, 8$. In the case of $\alpha =10$, the SIPM has a slightly different energy dissipation pattern from the other two numerical methods.

\subsection{Domain wall motion}
\begin{figure}[htbp]
	\centering
	\subfloat[Magnetization for initial state]{\label{NeelWall_initial_mag}\includegraphics[width=2.8in]{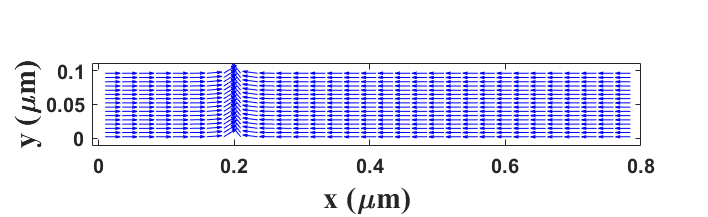}}
%	\hspace{0.1in}
%	\subfloat[Magnetization with $\alpha=2$ at $1\,$ns]{\label{NeelWall_alpha_2_1_mag}\includegraphics[width=2.8in]{NeelWall_alpha_2_FieldFixed_1ns_timescale_v1_mag_v1.png}}	
	\subfloat[Magnetization with $\alpha=2$  at $2\,$ns]{\label{NeelWall_alpha_2_2_mag}\includegraphics[width=2.8in]{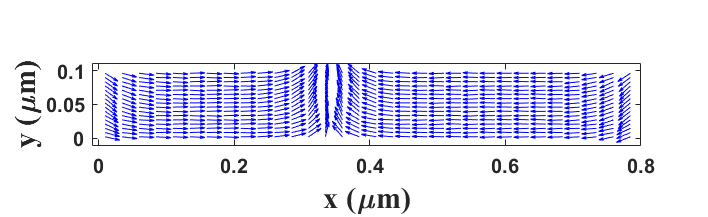}}
	\hspace{0.1in}
%	\subfloat[Magnetization with $\alpha=5$ at $1\,$ns]{\label{NeelWall_alpha_5_1_mag}\includegraphics[width=2.8in]{NeelWall_alpha_5_FieldFixed_1ns_timescale_v1_mag.png}}
	\subfloat[Magnetization with $\alpha=5$ at $2\,$ns]{\label{NeelWall_alpha_5_2_mag}\includegraphics[width=2.8in]{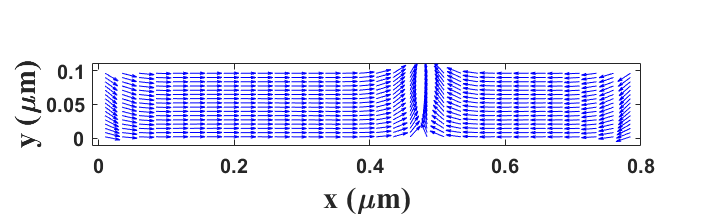}}
%	\hspace{0.1in}
%	\subfloat[Magnetization with $\alpha=8$ at $1\,$ns]{\label{NeelWall_alpha_8_1_mag}\includegraphics[width=2.8in]{NeelWall_alpha_8_FieldFixed_1ns_timescale_v1_mag.png}}
	\subfloat[Magnetization with $\alpha=8$ at $2\,$ns]{\label{NeelWall_alpha_8_2_mag}\includegraphics[width=2.8in]{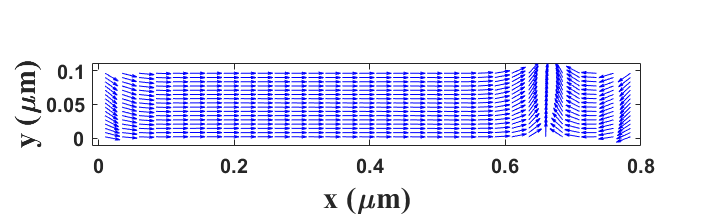}}
	\caption{Magnetization profiles of Ne\'{e}l wall motion in the presence of a magnetic field $\h_e=5\,$mT, with $\alpha = 2,5,8$ at $2\,$ns for the proposed numerical method. The in-plane arrow denotes the first two components of the magnetization vector. The wall moves faster for larger values of $\alpha$ and its velocity depends linearly on $\alpha$.}\label{NeelWall_alpha_2ns}
\end{figure}

\begin{figure}[htbp]
	\centering
	\subfloat{\label{velocity_alpha_varied_He}\includegraphics[width=2.8in]{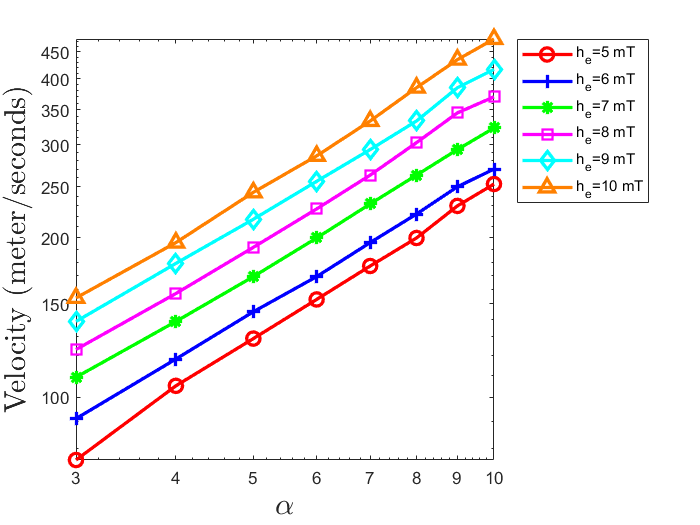}}
	\subfloat{\label{velocity_He_fixed_alpha}\includegraphics[width=2.8in]{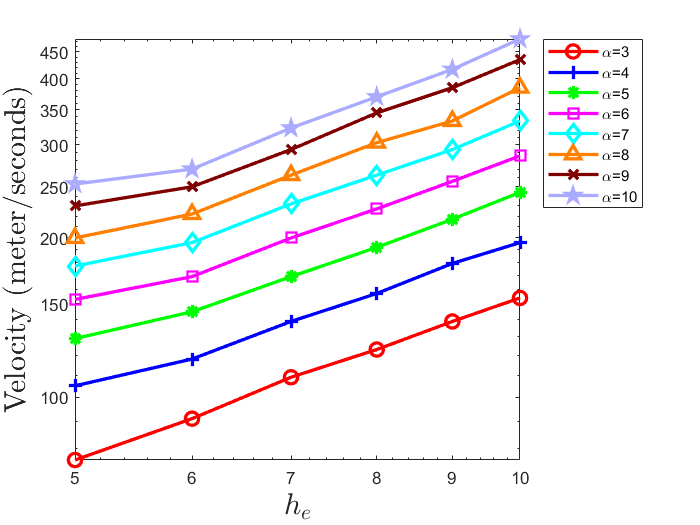}}
	\caption{Linear dependence of the wall velocity with respect to the damping parameter $\alpha$ (left) and the external magnetic field $\h_e$ (right).}\label{velocity_alpha_He}
\end{figure}

A Ne\'el wall is initialized in a nanostrip of size $800\times100\times4\,\textrm{nm}^3$ with grid points $128\times64\times4$. An external magnetic field of $\h_e=5\,$mT is then applied along the positive $x$ direction and the domain wall dynamics is simulated up to $2\,$ns with $\alpha=2,5,8$. The corresponding magnetization profiles are visualized in \cref{NeelWall_alpha_2ns}. Qualitatively, the domain wall moves faster as  the value of $\alpha$ increases. Quantitatively, the corresponding dependence is found to be linear; see \cref{velocity_alpha_He}. The slopes fitted by the least-squares method in terms of $\alpha$ and $\h_e$ are recorded in \cref{tab-3}.
\begin{table}[htbp]
	\centering
	{\caption{Linear dependence of the domain wall velocity $V$ in terms of the external magnetic field $\h_e$ and the damping parameter $\alpha$.} \label{tab-3} }{
		\begin{tabular}{c|c|c|c|c|c|c|c}
			\hline 
			\diagbox{$\alpha$}{$V$ (m/s)}{$\h_e(\textrm{mT})$}&5&6&7 &8 & 9&10 & Slope\\
			\hline
			3& 76 & 91 &109  & 123 & 139&154 &1.024\\
			4& 105 & 118 & 139 & 157 &179 & 196&0.928\\
			5& 129& 145& 169 & 192 & 217&244 &0.932\\
			6& 153 &169 & 200 & 227 &256 &286 &0.927 \\
			7& 177& 196& 232 & 263 & 294& 333& 0.927\\
			8& 200 & 222 & 263 & 303 &333 &385 &0.954 \\
			9& 230 & 250 & 294 & 345 &385 &435 &0.954 \\
			10& 253 & 270 & 323 & 370 &417 & 476&0.943 \\\hline
			Slope & 0.984 &0.910  &0.910  & 0.933 &0.917 & 0.950 &--\\
			\hline 
		\end{tabular}
	}
\end{table}

\section{Conclusions}
\label{sec:conclusions}

In this paper, we have proposed a second-order accurate numerical method to solve the Landau-Lifshitz-Gilbert equation with large damping parameters. For the numerical convenience, the LLG system is reformulated so that in which the damping term is rewritten as a harmonic mapping flow .This numerical scheme is based on the second-order backward-differentiation formula approximation for the temporal derivative, combined with an implicit treatment of the constant-coefficient diffusion term, and the fully explicit extrapolation approximation of the nonlinear terms, including the gyromagnetic term and the nonlinear part of the harmonic mapping flow. Thanks to the large damping parameter, the proposed method is verified to be unconditionally stable. %Precisely, the damping parameter needs to be greater than $3$ in order to obtain the second-order accuracy \cite{Xie2021EElinear}. 
The proposed method is much more efficient than other semi-implicit schemes since only symmetric, positive definite linear systems of equations with constant coefficients need to be solved. Meanwhile, the proposed method is more accurate than the standard Gauss-Seidel projection method, due to its second-order accuracy in time. Numerical results in 1D and 3D are provided to demonstrate the accuracy and the efficiency of the proposed numerical method. In addition, micromagnetics simulations using the proposed method have provided physically reasonable structures and captured the linear dependence of the domain wall velocity with respect to the damping parameter. Therefore, the proposed method could be efficiently used for challenging practical simulations of micromagnetics with large damping parameters.

\section*{Acknowledgments}
This work is supported in part by the grants NSFC 11971021 (J.~Chen), NSF DMS-2012669 (C.~Wang), NSFC 11771036 (Y.~Cai).

\bibliographystyle{amsplain}
\bibliography{references}

\end{document}